# The 2025 Roadmaps for the US Magnet Development Program

***Compiled by***
Lance Cooley, Paolo Ferracin, Steve Gourlay,
David Larbalestier, Mark Palmer, Soren Prestemon[*], George Velev

&

***With Major Contributions from Technical Leads***
Giorgio Ambrosio, Diego Arbelaez, Karie Badgley, Lucas Brouwer, Daniel Davis, Jose Luis Fernandez, Vadim Kashikhin, Steven Krave, Maxim Marchevsky, Igor Novitski, Ian Pong, Tengming Shen, Stoyan Stoynev, Reed Teyber, Giorgio Vallone, Xiaorong Wang, Xingchen Xu

and Collaborators
within the US MDP

---

[*] Corresponding author: soprestemon@lbl.gov

# Table of Contents

# Executive Summary

The US Physics community completed the Snowmass planning process in 2022, culminating in the HEPAP Particle Physics Project Prioritization Panel (P5) publishing its summary report at the end of 2023. Building on this, the US Magnet Development Program, a national accelerator magnet R&D program established by DOE-OHEP in 2016, has updated its strategic plan to align with the 2023 P5 report, resulting in this roadmap document.

The primary goals of the program have been modified to address two major considerations in the P5 report. First, the High Energy Physics community has identified a 10TeV parton Center of Mass (pCOM) collider as critical for Beyond the Standard Model (BSM) physics exploration. Second, there are significant synergies with other DOE programs and with industry that can be leveraged to more rapidly and efficiently develop critical magnet technology for HEP. Collider concepts being explored to address the need include a) a hadron (e.g. proton-proton) collider at the ~100TeV scale, b) a muon collider at the 10TeV scale, and c) advanced accelerator concepts using $e^+e^-$ collisions at the 10TeV scale. Concepts (a) and (b) both require advanced accelerator magnet technologies beyond the current state of the art; in additional, the muon collider requires advanced solenoid technology to enable the beam-cooling needed to achieve the luminosity required.

Advances by MDP over the last approximately 8 years have led to record accelerator dipole fields, the development and demonstration of stress-management concepts that pave the way to both high-field, large-bore dipoles and to hybrid accelerator magnets utilizing both low- and high-temperature superconductors, and the exploration of novel impregnation materials that hint at the potential for training-free magnet technology. Furthermore, the program has invested in the development of HTS magnet technology, using both Bi2212 and REBCO commercially available conductor architectures. The program is poised to integrate these developments in a suite of prototype magnets that explore these new regimes of high magnetic field that will provide critical performance insights to guide plans for hadron and muon colliders.

To support the broader challenges associated with a muon collider, the program now includes research on solenoid technology tailored to HEP needs – the intent here is to collaborate with experts in the field, leveraging the tremendous progress made in the field by the National High Magnetic Field Laboratory (NHMFL), but also the more recent investments in the field by the fusion community. Synergies abound – our research will focus on addressing HEP-specific aspects, but we anticipate significant mutual interest in collaborating to rapidly advance high field solenoid technology, in particular using HTS materials.

A major element of MDP since its inception is the investment in core technologies, including advanced modeling, the development of unique diagnostics and instrumentation, the exploration of novel means to enhance training rate in LTS magnets, and materials research to support industry advances in conductor performance. These investments continue, with increased focus on integrating advances into magnet prototypes. As an example, for HTS magnets we expect diagnostics and instrumentation to evolve from tools of discovery and understanding, to tools integral to the magnet operation and protection.

To further support HEP initiatives to explore 10TeV pCOM colliders, we have added program elements to interface with anticipated future accelerator design studies; interface with interaction-region studies, where high-field accelerator magnets are critical; and to explore higher temperature operation of magnets, with particular attention to the impact on sustainability, e.g. in terms of operation power consumption.

Finally, in light of the critical importance of having sufficient expertise trained and available to embark on a future collider, the MDP program now explicitly recognizes workforce development as part of its mission. MDP has a proven record of attracting and developing the next generation of talent, and we intend to further strengthen that role in support of HEPs future needs.

# Introduction

Today's colliders are built on a foundation of superconducting magnet technology that provides strong dipole magnets to maintain the beam orbit and strong quadrupole focusing magnets to achieve the extraordinary luminosity required to probe physics at the energy frontier.

The US Magnet Development Program (MDP) was initiated in late 2015, following recommendations from the 2013 P5 report and the follow-on report from the HEPAP Accelerator Subpanel. MDP brings together teams from the leading US accelerator magnet research programs to develop the next generation of magnet technology for future collider applications, with primary members being Brookhaven National Laboratory (BNL), Fermi National Accelerator Laboratory (FNAL), Lawrence Berkeley National Laboratory (LBNL), and the National High Magnetic Field Laboratory (NHMFL). Sponsored by the DOE Office of High Energy Physics, the program strives to maintain and strengthen US Leadership in the field, while nurturing cross-cutting activities from other programs to further strengthen the research and its impact to the DOE Office of Science. General long range magnet R&D is a critical foundation for HEP, providing a strong basis to develop strategies for future colliders, growing the next generation of scientists and engineers in the field of accelerator magnets, so as to provide the foundation for directed-R&D that will ultimately be required to scale up and finalize magnet designs for a future collider project (see Figure 1). The vision and the goals of the original US-MDP have been reviewed and slightly modified, to fully align with the 2023 P5 report, and in particular to increase focus on the need to nurture and develop the workforce expertise needed to deliver the next generation collider.

### Vision:
Enable new energy-frontier colliders to probe physics beyond the Standard Model

### Mission:
Expand US leadership in high-field accelerator magnet technology to enable the next generation of High Energy Physics colliders

### Goals:
- Explore and define the performance limits of superconducting accelerator magnets
- Develop, understand and demonstrate high field HTS magnet technology
- Investigate and understand the fundamental science of magnet design and performance
- Pursue conductor R&D to achieve properties that align with accelerator magnet goals
- Support the development of advanced workforce for superconducting magnet technology

Furthermore, MDP has the goals of:
- Integrating the teams across the partner laboratories for maximum value and effectiveness to the program
- Identifying and nurturing HEP and cross-cutting / synergistic activities and opportunities with other programs to more rapidly advance progress towards our common goals.

*Figure 1 The US DOE approach balances long-range R&D and project preparation.*

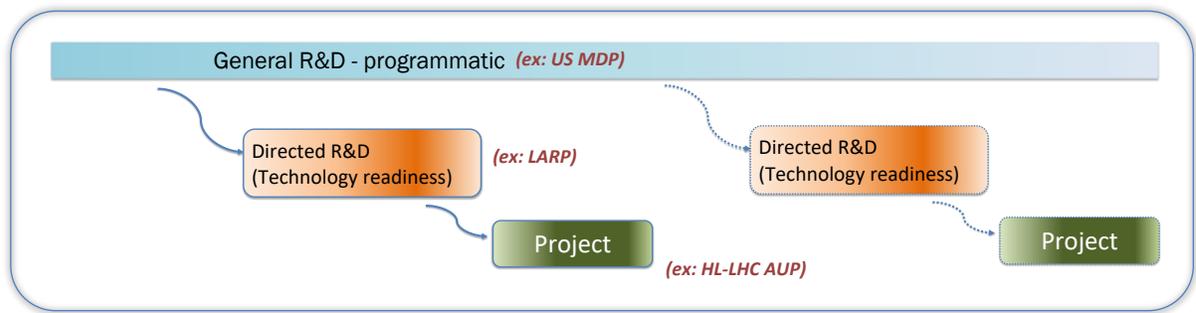

## Program Overview and Alignment with HEP Strategy

The mission of the US Magnet Development Program (MDP) is to develop advanced superconducting accelerator magnet technology for future HEP colliders. The physics motivation for colliders has been developed by the US and international communities through long-range planning processes, including the European Strategy update process and the decadal Snowmass and P5 process in the US. Researchers from MDP are strongly engaged in the international community and contributed significantly to the 2021 Snowmass process and the 2023 P5 report [1] guiding HEP strategic planning and investments.

The Snowmass process identified 10TeV parton Center of Mass (pCOM) colliders as the means to provide the most comprehensive increase in discovery potential for physics beyond the Standard Model. Two of the primary candidates for such a collider are a hadron collider (e.g. the FCC-hh) and a muon collider; both require advances in superconducting magnet technology to deliver on the physics promise. While the MDP has since its inception worked to address the needs of a future hadron collider, the renewed interest in a muon collider as a viable alternative for the energy frontier has motivated some realignment of the MDP research structure in this update to our roadmap. Furthermore, in particular for a muon collider, there are significant synergies with other DOE programs and with industry that can be leveraged to more rapidly and efficiently develop critical magnet technology for HEP (see Figure 2 and section *Synergistic Programs and Activities*).

The properties of the superconductors used in high field accelerator magnets are critical to magnet performance and are a major element in the magnet cost. The ability to continue to innovate new superconductor architectures, and to bring new conductors to industrial maturity for projects, is a central activity of magnet development. Nurturing and maintaining a vibrant, competitive ecosystem for superconductor development as well as for reliable, high quality production capability, will require long term strategic planning and commitment from DOE, in close collaboration with industry. MDP continues to support this through targeted investments in superconductor (see section *Conductor Procurement and Research and Development*).

> *P5 report:* **Exploring the Quantum Universe: Pathways to Innovation and Discovery in Particle Physics** [arXiv:2407.19176](arXiv:2407.19176)
>
> *"We note that there are many synergies between muon and proton colliders, especially in the area of development of high-field magnets. R&D efforts in the next 5-year timescale will define the scope of test facilities for later in the decade, paving the way for initiating demonstrator facilities within a 10-year timescale (Recommendation 6)."*

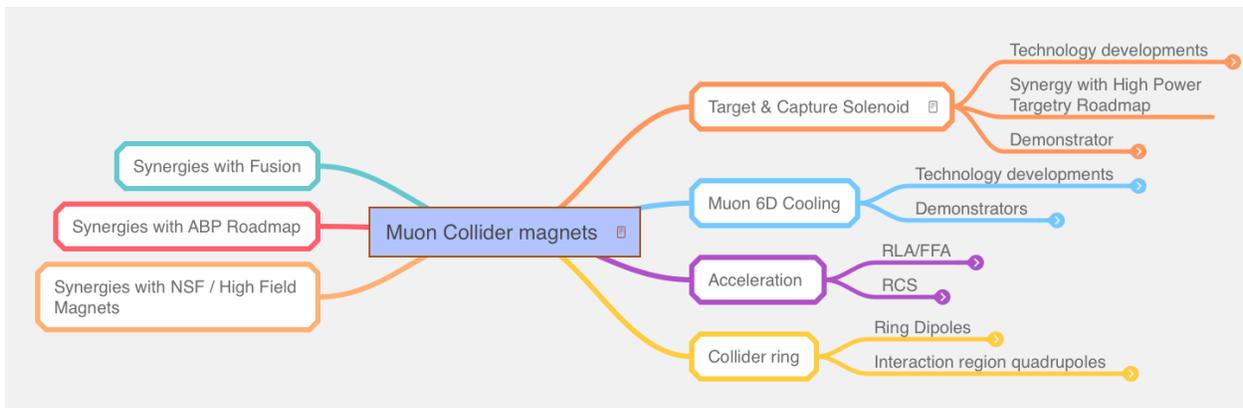

*Figure 2. A schematic of magnet-related research topics for muon colliders (to the right) and associated synergistic activities (to the left).*

## Program Structure

The revised MDP program structure (Figure 3) is designed to advance and demonstrate high field magnet technology for high energy physics, with a strong component of its research in supporting areas that build our scientific understanding of magnet behavior so as to enable more cost-effective, efficient, and higher performing magnets in the future. The 2023 P5 report focused on 10TeV pCOM colliders as major future initiatives to explore the energy frontier; our program therefore has an additional element of forward-looking exploratory studies designed to support these initiatives with insights into performance limitations, sustainability considerations, and possible paradigm shifts in collider design.

*Figure 3. Program structure with four dedicated areas.*

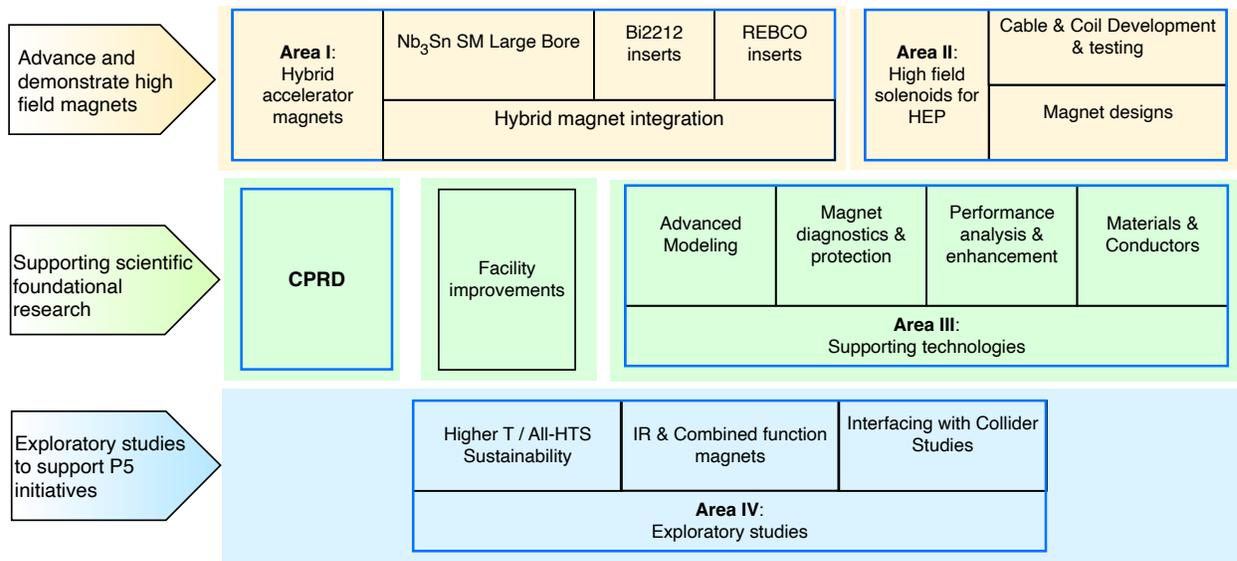

## Advancing and Demonstrating High Field Magnets

At the core of MDP is the development and demonstration of high field accelerator magnets for high energy physics. Since its inception, MDP has worked to develop the magnet technology to enable colliders with new reach, exemplified for example with the 14.5T record field achieved by the four-layer cosine theta magnet produced by MDP in 2019 [2], and in the development of stress-managed magnet concepts such as the "Canted Cosine-Theta" (CCT) and the "Stress-Managed Cosine Theta" (SMCT) designs [3, 4]. In these designs the Lorentz forces are captured by mechanical structures locally, rather than allowed to accumulate on the magnet midplane as is the case with traditional "Cosine-Theta" designs used in colliders to-date.

Beyond requiring advances in accelerator magnet technology such as that enabled by MDP research to-date, a future muon a collider will also need advances in solenoid magnet technology, and the new MDP roadmap includes dedicated research in that arena, leveraging strong synergies with ongoing efforts at the NHMFL and in the US fusion community. MDP Areas I and II are designed to support these two critical magnet development needs.

### Area I: Hybrid Accelerator Magnets

Using HTS-based dipoles within the bore of a $Nb_3Sn$ dipole magnet is anticipated to be the most cost-effective means of pushing HTS accelerator magnet technology to higher fields. To-date, MDP has focused on developing the core elements of a hybrid magnet: a) stress-managed, large-bore $Nb_3Sn$ magnet technology, exemplified by the SMCT and CCT designs; and b) HTS accelerator magnet technology with REBCO and Bi2212 cables. Our goal now is to integrate these - up until now separate - thrusts into working hybrid accelerator magnets. Thus, we have a twofold ambition: first, to bring the LTS SMCT and CCT magnets to fruition as exceptional demonstrators of 12-14T large-bore dipoles relevant to applications such as a muon collider and as workhorse magnets for developing and demonstrating practical LTS/HTS hybrid magnet technology; and second, to further demonstrate HTS accelerator magnet technology in high magnetic fields.

> **Driving questions:**
> - Can stress management be used to produce high field magnets with large apertures that can approach the conductor limits without degradation?
> - What are the most effective impregnation materials regarding training, degradation, radiation hardness, etc?
> - Stronger support materials are needed for fields > ~16 T magnets. How does this affect fabrication of components (machinability, scale up, 3D printing, etc.)?
> - Can training be significantly reduced / eliminated and degradation avoided by using stress management together with wax or Telene-based impregnation for large aperture (> ~ 120mm) high field magnets (>~10T)?
> - Are HTS conductors compatible with the hybrid HTS/LTS approach and what needs to be done to improve the compatibility?
> - How do we minimize critical current degradation in HTS materials during coil and magnet fabrication, under Lorentz forces, and due to thermo-cycling?

### Sub-area I.a: $Nb_3Sn$ Stress-Managed Large Bore Magnets

The development of stress managed large bore magnets is central to MDPs mission. The large bore challenges conventional cos-theta magnet designs at high field, and hence is an excellent proving ground for the concept of stress-management. Successful demonstration of reliable large bore ($\geqq$120 mm) dipoles in the 12-14T range, without exhibiting degradation due to strain and under operational cycling, is a major milestone for the program. Furthermore, the demonstration will serve as a critical gate for a Muon collider, where high field, large bore dipoles are required to achieve high luminosity while accommodating significant radiation shielding in the bore. Finally, such large bore dipoles will serve as workhorse magnets for the hybrid magnet program, enabling development and testing of HTS dipoles in high magnetic fields.

### Sub-area I.b: Bi2212 insert dipoles

The DOE-OHEP has invested significantly in Bi2212 - an isotropic, round-wire conductor with built-in current sharing - over the last decade. Bi2212 performance continues to evolve based on deeper understanding of the conductors, which has encouraged and enabled new powder suppliers and wire manufacturers to appear. The accelerator dipole technology has advanced, with multiple magnets demonstrating progress on key challenges, e.g. insulation, overall $J_c$, over-pressure reaction, etc. that encourage advancement to the next stages. MDP is poised to perform the first hybrid test using a Bi2212 insert in a 90mm bore Nb$_3$Sn magnet, and coils have been fabricated in preparation for testing in the flagship 12-14T Nb3Sn outsert magnets, with 120mm-bore dipoles, currently under development. The program will now focus on the development and testing of Bi2212 dipoles in high-field background magnets, where challenges of quench protection and strain management will need to be addressed.

### Sub-area I.c: REBCO insert dipoles

The superconductor REBCO has unique characteristics and offers the highest field and temperature performance potential of any superconducting material. The tape architecture, combined with the intrinsic superconducting anisotropy of the material, introduces distinct challenges for both magnet design and manufacture as well as for magnet protection. The opportunity, however, is tremendous due to the strong pull currently underway from the fusion community, where the promise of high-field, compact fusion devices is fueling extraordinary investments in REBCO conductor development.

Our future program will leverage recent MDP developments to focus on cable and magnet design architectures that are compatible with insert magnet geometry constraints while probing REBCO dipole magnet operation in high field. Focus will be placed on addressing potential strain degradation that may occur during magnet fabrication and operation, as well as on developing and demonstrating reliable magnet protection and quench mitigation strategies.

### *Area II: High Field Solenoids for HEP*

The high energy physics community has identified a number of opportunities to explore new physics that exploit colliders and novel experiments, many of which require advanced high field solenoids in addition to the HEP "workhorse" bending dipoles and focusing quadrupoles. Prime examples include a muon collider and various axion search experiments. Major advances in high field solenoids over the last decade have been led by the National High Magnetic Field Laboratory, motivated primarily by the provision of >30 T magnets for NHMFL users, and culminating most recently in the all-superconducting hybrid LTS/REBCO 32 T magnet. The 2023 P5 report, coupled with the 2024 NAS report [5], motivate a strong partnership between MDP and the NHMFL to further advance high field solenoid technology tailored to the needs of HEP. This is a new Area within MDP, and the goal is to leverage synergies with the NHMFL and Fusion applications to rapidly develop solenoid technologies addressing HEP needs.

### Sub-area II.a: Cable and Coil Development and Testing

A unique characteristic of HEP applications utilizing solenoids is the diversity of scale of their usage. For muon colliders, solenoid needs range from a few unique, very large bore, high field solenoids for the muon production area, to a kilometer-scale suite of high-field solenoids for the muon-cooling section. For axion searches, a range of field strength and bore sizes are of interest, but generally the search sensitivity scales linearly with field volume and quadratically with field strength, motivating large bore, high field solenoids (we note that some axion search concepts utilize dipoles, with similar field and bore scaling). Together, these applications motivate the use of scalable conductor architectures, allowing magnet designers to tailor the inductance to optimally address powering and protection considerations in light of the extraordinary amount of magnetic stored energy in the systems. Although scalable conductor architectures are routinely used in the large solenoids of collider detectors, they have not been incorporated in high field NMR, with the (perhaps unique) exception of cable-in-conduit conductor (CICC) usage in the Series-Connected Hybrid magnet currently in operation at the NHMFL. Such architectures are, however, under development by the nascent - but rapidly growing - compact fusion community.

We intend to focus MDP's first experimental efforts in the solenoid field in exploring and developing cable architectures - primarily using HTS materials - in a high-field solenoid background. Major goals are to determine performance boundaries and to develop quench protection methodologies that can be scaled.

Sub-area II.b: Magnet Designs

The unique characteristics of HEP solenoid needs, and in particular the likely use of scalable conductor architectures, motivate a review and adaptation of solenoid magnet design with respect to the high-field solenoid designs that are typically used in high-field solenoid user facilities. Managing the large hoop, radial, and axial stresses in a reliable and cost-effective manner, and designing magnet protection methodologies that are compatible with the scale of field and volume envisioned, are significant challenges that require in depth design, prototype construction, and analysis work. Due to the scale of the magnets envisioned, building demonstrations of such magnets is beyond the budget constraints of MDP; this is an area where synergies with other programs may be enabling. What is within MDP capabilities is to build high stress, high field demonstrators that take HTS conductors, still very primitive and evolving rapidly, into domains of field and stress of interest to future HEP uses, so as to identify and retire "unknown unknown" risks and to set up project driven prototypes.

## Supporting Foundational Science to Advance Magnet Technology

Lasting progress in advancing magnet technologies relies on a thorough understanding of the underlying physics mechanisms and the technical elements that enable and control them. This understanding is rooted in experiments, accompanied by instrumentation and diagnostics, that test and challenge our expectations, as well as models and simulations that predict magnet behavior based on our knowledge. The agreement between experimental results and simulations serves as a key indicator of our understanding.

*Area III: Supporting Technologies*

To provide focus and alignment with MDP goals, four technology arenas have been identified that are critical to support the advance of magnet technology. These technological underpinnings are at the heart of MDP research (see Figure 4) - enabling breakthroughs in magnet performance, providing an understanding of magnet behavior leading to reliable operation, guiding paths to scalable designs that can adapted to industrialization, and opening avenues towards more sustainable colliders. The supporting technology areas are therefore challenged to develop and innovate new ideas while simultaneously contributing directly to the design, fabrication, and testing of magnets in Areas I and II.

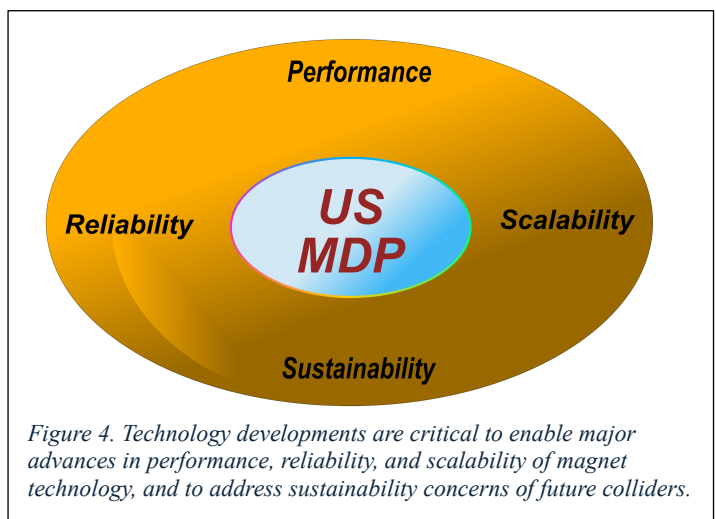

*Figure 4. Technology developments are critical to enable major advances in performance, reliability, and scalability of magnet technology, and to address sustainability concerns of future colliders.*

### Sub-area III.a: Advanced Modeling

The ability to accurately predict magnet performance using modeling techniques is an essential ingredient to demonstrate understanding of the physics and engineering that drive the performance. Most importantly, the models can guide further magnet technology advances, and support the development of design and fabrication specifications and tolerances for applications. Due to the complex interplay between mechanical, electromagnetic, and thermal phenomena, and to the vast range of physical scales involved, a variety of advanced modeling techniques are needed that incorporate multi-physics and that can communicate between each other. To the degree possible, we strive for open-source models to allow broad community usage and to encourage further improvements from collaborators.

The Advanced Modeling effort aims to leverage state-of-the-art computational tools and methodologies to enhance the design and performance of superconducting magnets. This effort will be focused on three primary areas: development of design tools, fundamental understanding of conductor, cable, and magnet behavior, and performance limitation studies. Each area contains main efforts aligned with the needs and challenges of the other R&D areas.

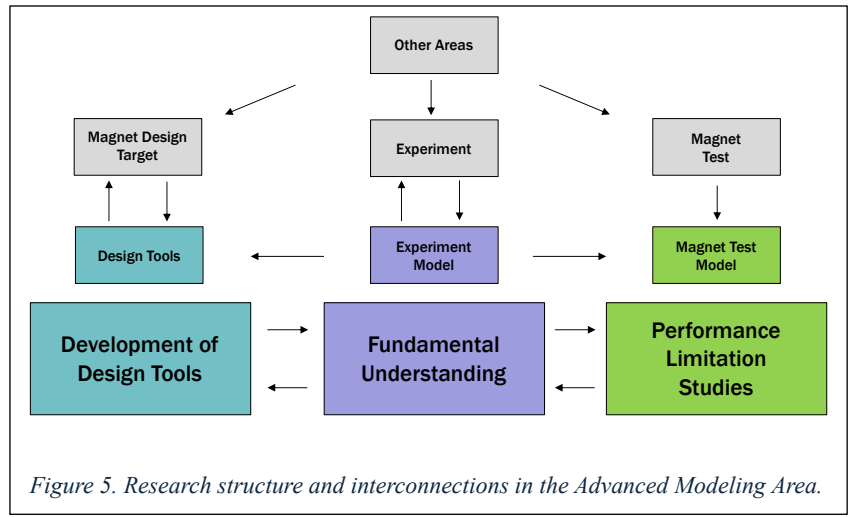

*Figure 5. Research structure and interconnections in the Advanced Modeling Area.*

### Sub-area III.b: Magnet Diagnostics and Protection

Diagnostics and instrumentation are the basis for all magnet performance information obtained during magnet fabrication and testing, and are therefore critical to the MDP mission. Multi-sensor diagnostics data obtained in the R&D magnet tests, combined with AI/ML processing and advanced models, enhance our understanding of magnet performance and inform the path to further technological advances.

Uncovering physical mechanisms responsible for premature quenching and training is critical for overcoming performance degradation and training in high-field LTS magnets that operate close to the conductor's mechanical stress limits. New diagnostics must be developed to probe mechanical energy conversion into heat in these critical conditions. Well-controlled small-scale experiments can be especially useful in that respect, helping us better understand the transient thermo-mechanical phenomena and guiding our search for new impregnation materials and techniques for future record-field LTS magnets.

For HTS magnets, the isotropic, multifilament, good-current-sharing nature of Bi-2212 eases the design methodology which for anisotropic, single-filament REBCO tapes becomes much more challenging. Developing techniques allowing real-time detection and localization of hot spots, especially in REBCO magnets, is of major importance. Non-voltage-based techniques appear promising, especially those based on sensing localized temperature increases within the magnet windings or current re-distribution between superconducting cable elements. Different sensing modalities (ultrasonic, RF-based, Hall sensors, fiber-optics) recently explored by the MDP program must now be scaled up and integrated into prototype magnet coils to compare their efficiency. In connection with this effort, we will also develop diagnostic instrumentation for in-situ localization of HTS conductor defects in magnet coils and quantifying current sharing in complex conductors.

Quench protection is vital for safe and reliable magnet operation. We note that it fully relies on the early warning signals produced by the quench detection system. Various novel quench detection and protection techniques have been proposed and tested recently, including active current control with cryogenic power electronics, coupling-losses induced heating (CLIQ), smart insulation-based control of current sharing, etc. Variations and combinations of those techniques must be explored further to protect future HTS and hybrid high-field magnets of very large stored energy. Specifically, the interaction between LTS and HTS protection systems is a critical R&D topic for the hybrids. As HTS accelerator magnet technology matures, we expect magnet diagnostics and instrumentation to become fully integrated into magnet design, fabrication, and testing. The ultimate goal of this effort is to provide a robust, reliable, and self-consistent quench detection and protection solution for both stand-alone HTS and LTS/HTS hybrid magnets.

Sub-area III.c: Performance Analysis and Enhancement

Data from magnet testing is providing an ever-expanding foundation for analyzing performance correlations as a function of a wide range of factors, including design parameters and coil and magnet training history. Systematically analyzing this data, along with identifying and conducting specific test campaigns to explore causal connections, may yield deeper insights into the factors driving magnet performance and guide future magnet development. Again, well-controlled small-scale experiments and targeted developments have the potential to gradually enhance our understanding of key phenomena and aid in building and validating computer models of magnet behavior. These experiments, combined with new techniques and approaches, will lead to faster and more sustainable progress in magnet performance for both LTS and HTS. The cumulative impact of these efforts shapes a sub-area that is closely linked to many other sub-areas, with bidirectional information flow being crucial for advancements across all fields.

Sub-area III.d: Materials and Conductors

All superconductors beyond Nb-Ti are strain sensitive and brittle in nature; in the case of $Nb_3Sn$ and Bi2212, the conductors are subjected to complex heat-treatments to form the superconducting compounds, and in most cases for accelerator magnets the heat treatments are performed after the coil is wound (i.e. "wind and react" technology) due to the strain sensitivity of the superconductor after heat treatment and the realities of the magnet geomtry. To advance accelerator magnet technology, it is essential that we have a thorough understanding of the material properties at all stages of the fabrication process, and that associated magnet materials be selected and applied in a manner that does not degrade magnet performance. Developing a robust database of materials properties and a process for materials selection and usage is a critical element of MDP.

*Facilities Improvements*

A suite of state-of-the-art facilities is essential to support the magnet tests associated with the Program described above. Short-sample testing capabilities exist in multiple laboratories; improvements are needed to meet the increasing demands of new state-of-the-art $Nb_3Sn$ superconductors as well as HTS conductors. Facilities such as the 10T Common Coil at BNL that enables testing of cables and coils in unique field and stress environments will require enhancements to enable a broader spectrum of experiments for MDP. Most critically, the ability to test hybrid magnets is an essential component of the program. This includes separately powered outsert and insert magnets, each requiring unique and flexible quench detection systems. Moreover, power switching and fast energy extraction capabilities are crucial, especially for the protection of the HTS component in hybrid magnets.

Additionally, advanced magnet diagnostics and instrumentation demand specialized data acquisition capabilities, and the test facilities must evolve in step with MDP technology developments. Identifying new enabling technologies, implementing best practices, and sharing magnet test experiences and data are all vital for the success of the MDP mission.

*Conductor Procurement and Research and Development*

The Magnet Development Program requires access to state-of-the-art superconductors to enable rapid and successive development and testing of accelerator magnets. Furthermore, superconductor performance drives accelerator magnet performance, motivating continued feedback from MDP to the superconductor industry to enable conductor advances that benefit HEP. The mission of Conductor Procurement and R&D (CPRD) is therefore a) to identify future MDP magnet conductor needs and to procure the requisite conductor in a timely manner, and b) to work with industry to identify possible research opportunities that may lead to significant advances in superconductor performance. These R&D investments should complement existing funding opportunities such as those possessed by University partners of MDP and SBIR, and should support the competitiveness of the US superconductor industry.

## Exploratory Studies

As mentioned above, the 2023 HEP P5 report identified a number of potential collider opportunities over the coming decades that will require advanced accelerator magnet technology to come to fruition. The facilities envisioned are of an extraordinary scale that will require international collaboration beyond the current level and the support from the broader public across many nations. Sustainability – both in the facility construction and in its long-term operation – will be an essential ingredient in getting humanity behind such an endeavor. To optimally prepare for such colliders, it is imperative that we collaborate closely with the accelerator design initiatives and identify the magnet advances that provide optimal value to the accelerator performance [6]. Tradeoffs between field, field quality, aperture, operating temperature, magnet cost, and operational power requirements, for example, may play a critical role in the design of future facilities.

*Sub-Area IV.a: Higher Temperature / All HTS and Sustainability*

Superconductors enable modern colliders by providing the beam's guiding magnetic field without energy loss, i.e. the magnetic field is effectively stored potential energy. Nevertheless, there is energy consumed in the refrigeration power needed to keep the magnets at their operating temperature. The thermodynamic optimal, Carnot efficiency, scales with $Q_c \sim T_{warm}/T_{cold}$, e.g. increasing the operating temperature from 1.9 K to 4.2 K results in more than a factor 2 reduction in wall-plug power. Operating at 20 K would improve Carnot efficiency by a further factor of nearly 5. However, operating at higher temperature limits superconductor options, and comes at a cost of more superconductor due to the *Jc(T)* dependence. Furthermore, there are complex interrelations, for example related to superconductor magnetization and magnet protection, that can impact the viability of operation at higher temperature. Understanding the field and temperature limitations of each superconductor and its associated accelerator magnet technology, as well as the potential implications for collider sustainability, is a critical focus for MDP.

*Sub-Area IV.b: Interaction Region and Combined Function Magnets*

While future collider magnet technology typically focuses on ring dipoles due to their significant impact on overall facility cost and performance, additional accelerator magnets also play a crucial role in driving collider performance. Notably, the strong quadrupoles in the detector interaction region are essential for maximizing collider luminosity. Advances in accelerator magnet technology must address the unique challenges posed by these magnets.

Furthermore, innovative designs, such as combined-function magnets that enable unique lattice configurations, could introduce new collider design paradigms, ultimately enhancing overall physics capabilities.

*Sub-Area IV.c: Interfacing with Collider Studies*

Based on the 2023 HEP P5 report, we anticipate the formation of collider design group(s) within the next 2-3 years. MDP intends to encourage close collaboration and communication between those groups and MDP researchers. Identifying the primary limitations on magnet performance, e.g. through guidance on achievable fields, field quality, operating parameters, and other magnet considerations will be important contributions from MDP to enable credible collider designs. Furthermore, close communication can guide MDP developments on those elements of magnet advances that have most impact on collider performance, and/or that address the largest risks in collider operations.

Moreover, this sub-area will review ongoing design efforts and gather future requirements from the field to determine how MDP can contribute to magnet design, particularly for future interaction region magnets and combined-function magnets. In addition, we plan to act as liaisons to the broader community, suggesting how MDP can benefit accelerator magnet development in the US.

# Roadmaps for the US-MDP

A five-year roadmap detailing the major research thrusts of the program is shown in Figure 6. Major milestones for the program are provided in Appendix I. A master list of all milestones for the program is maintained as a "living document", with Areas providing technical progress updates on a regular basis; obsolete milestones can then be removed, and new developments may lead to the introduction of new milestones. The annual MDP collaboration meeting provides an avenue for formal review of all program milestones.

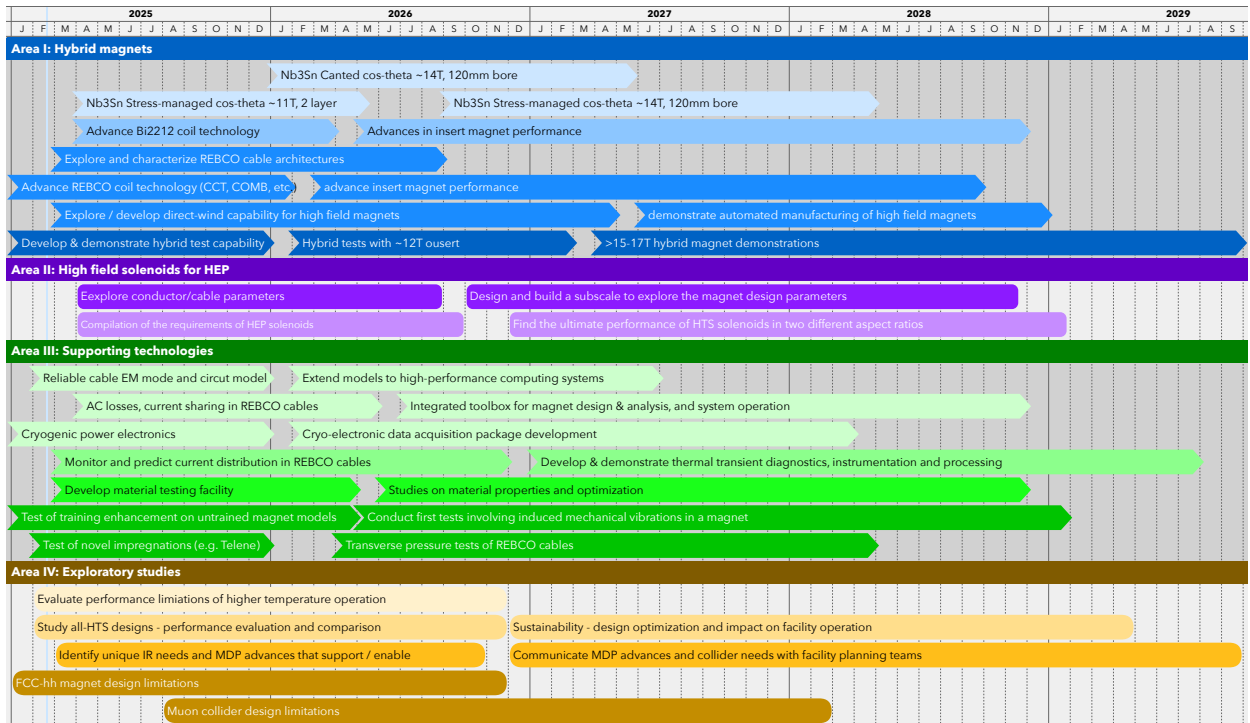

Figure 6. Roadmap for the program for the 2025-2030 period, aligned with the research Areas described above.

# Synergistic Programs and Activities

The landscape for superconducting magnet technology has evolved significantly over the last decade, most importantly with the nascent "compact fusion" industry and its investments in REBCO superconductors and related magnet technology. MDP research in HTS magnet technology, and in particular on the foundational science aspects (see Area III above), are synergistic and can support risk mitigation for industry, for example through the development of improved numerical modeling of REBCO tapes, cables, and magnets, and through the development of diagnostics and magnet protection paradigms for high-field, large-stored-energy REBCO-based magnet systems.

**NASEM Report: The Current Status and Future Direction of High-Magnetic-Field Science and Technology in the United States (2024) DOI 10.17226/27830**

*"A concerted effort from multiple agencies will bring both scientific and commercial advantages for the United States. Without higher magnetic fields, we will not have compact fusion devices, next-generation MRI, high-field nuclear magnetic resonance (NMR), or the muon collider."*

The renewed interest in muon colliders motivates MDP research into muon-cooling-relevant solenoid configurations; strong synergy exists with the ongoing advances in HTS solenoid technology spearheaded by the NHMFL in support of its user community, as well as solenoid requirements of compact Tokamak fusion machines. These, and other, synergies were identified in the recent National Academies report [5].

Beyond the HEP and FES interest in HTS magnet technology, we note the benefits of MDP research on other DOE Office of Science endeavors – examples include axion searches as strong candidates for dark matter, typically requiring strong magnetic fields; interaction region (IR) superconducting magnets for the Electron Ion Collider; electron cyclotron resonance (ECR) superconducting magnet systems for facilities such as FRIB; and superconducting undulators for light sources including storage rings and free electron lasers. These applications all push the frontiers of magnet technology and enable new science capabilities, and benefit from MDP research developments - a testament to the commitment of HEP to advance accelerator research for the DOE Office of Science.

## Collaborations

High Energy Physics is fundamentally an international endeavor, and the development of critical technologies enabling new colliders is similarly advanced through the combined efforts of international scale. All areas of the MDP benefit from close communication and collaboration with magnet and conductor R&D partners outside the US. Well-established communication channels ensure that we maximize progress through programs that are both competitive but also complementary.

Ties with international laboratories and universities have been, and we expect will continue to be, a critical element of our program. The US MDP leadership strives to foster a diverse, inclusive, and collegial culture that motivates innovation and open communication. Transparency in our purpose and research approach has proven to be effective in supporting the development of strong collaborations with international partners. As an example, MDP interacts closely with the European High Field Magnet program (HFM), with both complementary and synergistic program elements and benefitting mutually from expertise and facilities supporting each program's goals.

MDP maintains a "living document" of active collaboration activities, identifying scientific scope and points of contact at each collaborating institute, so each collaborating member recognizes both the value of the activity to their program, and the commitment they are making to the activity (see Appendix II). The activities are reviewed regularly by the respective programs.

> **2020 UPDATE OF THE EUROPEAN STRATEGY FOR PARTICLE PHYSICS**
> **Technical Report CERN-ESU-015**
>
> *"The particle physics community should ramp up its R&D effort focused on advanced accelerator technologies, in particular that for high-field superconducting magnets, including high-temperature superconductors."*

We note that a significant fraction of MDP staff were educated, wholly or partially, at international institutions. Exchanges of ideas, concepts, and results with international collaborators are central to our approach to research.

## Summary

The US Magnet Development Program is a mature national program sponsored by the DOE-OHEP designed to advance accelerator magnet technology to enable the next generation of High Energy Physics colliders. This new roadmap aligns the program with the recent 2023 P5 report and with DOE's strategic plan for High Energy Physics. In particular, the roadmap targets magnet needs for future 10TeV pCoM collider options. The combination of a focused research program leveraging expertise from major laboratories and universities in the field furthermore provides an outstanding basis for workforce development to address the future needs of DOE-OHEP.

As noted above, High Energy Physics is an international endeavor, and MDP has established strong collaborations and connections with laboratories, Universities, and industries around the globe.

Furthermore, there are strong synergies between MDP goals and the needs and aspirations in a wide range of applications, from DOE-SC offices such as Nuclear Physics, Fusion Energy Sciences, and Basic Energy Sciences, to industrial applications such as medical therapy and – most notably in recent years – fusion.

The next decade is expected to bring exciting and groundbreaking developments, including a decision on a future Higgs factory, potential beyond the Standard Model physics results from the HL-LHC, and improved understanding of both the physics motivation for a future energy-frontier collider and of the major technological challenges that must be addressed. MDP is working hard to address the technology challenges in the magnet arena and to pave the way for achievable – and hopefully affordable – solutions that will enable the next revolution in High Energy Physis. In parallel, the fusion community is expected to achieve major milestones, from first operation of ITER to first prototyping of a Tokamak by industry. Together, the magnet community is poised to enable major developments for society, and MDP is well-positioned to play a central role in those developments.

# Appendix I: Milestone table

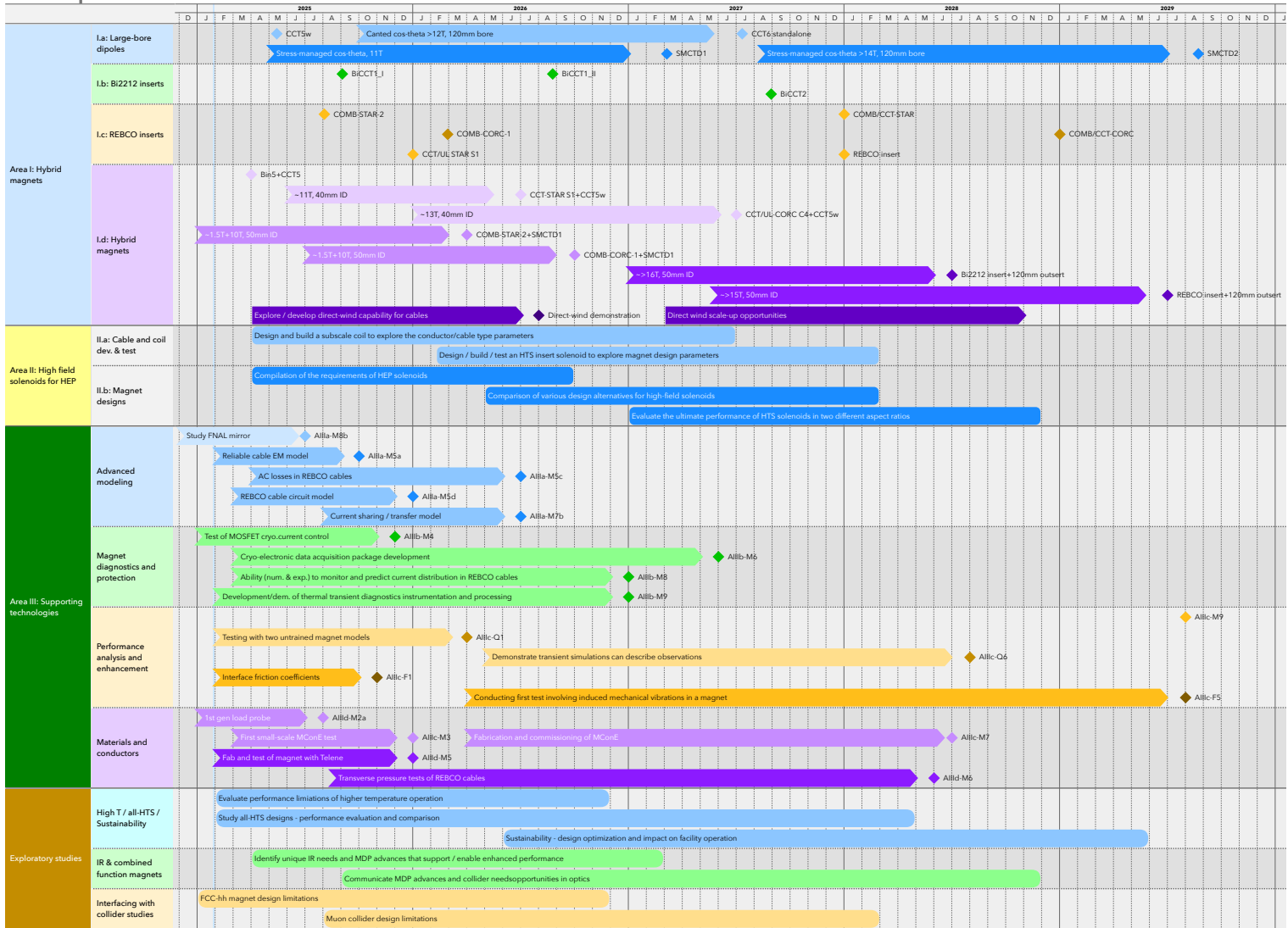

## Appendix II: Collaborations table

The Collaborations table is a "living" document that is maintained by the MDP Management team and updated on a bi-annual basis.